\documentstyle[12pt] {article}
\begin {document}

\begin{flushright}
{\bf US--FT/7--95}\\
\end{flushright}
\vskip 1.0cm

\begin{center}
{\Large {\bf BEAUTY HADRON PRODUCTION AT HERA--B}}\\
\vskip 0.3cm
{\Large {\bf ENERGY
IN DUAL PARTON MODELS}}\\
\vskip 1.5 truecm
{\bf N. Armesto, C. Pajares and Yu. M. Shabelski$^\ast$}\\
\vskip 0.5 truecm
{\it Departamento de F\'{\i}sica de Part\'{\i}culas,
Universidade de Santiago de Compostela,\\
15706--Santiago de Compostela, Spain}\\
\end{center}
\vskip 1.5 truecm
\begin{center}
{\bf ABSTRACT}
\end{center}

\vskip 0.3 truecm
Production of charmed and beauty mesons and baryons in hadron--nucleon
collisions at not very high energy is considered in the framework of the
Quark--Gluon String Model. We take into account the possible corrections
to the model
which disappear at asymptotically high energies but could be important at
comparatively low energies, essentially in the case of heavy flavour
production. The case of nuclear targets is also considered and the
$A$--dependence
of charmed and beauty hadron yields is calculated. Predictions for
the case of beauty hadron production at $\sqrt{s}$ = 39 GeV are
presented.

\vskip 4.0 truecm
\noindent $^\ast$Permanent address: Leningrad Nuclear Physics Institute,
Gatchina, Sanct--Petersburg 188350, USSR.\\

\vskip 0.7cm
\noindent{\bf May 1995}\\
{\bf US--FT/7--95}

\pagebreak
\noindent{\bf 1. INTRODUCTION}
\vskip 0.5 truecm
Heavy flavour production processes at high energies are considered usually
in the framework of QCD, see e.g. [1--3], which allows one to describe
quantitatively the cross sections of $c$, $b$ and $t$ quark production, their
dependence on the transverse momentum of the heavy quark, the rapidity
distributions, etc.
However,
these approaches do not allow to calculate the spectra of different mesons
and baryons. In the case of high energies there is also the problem of the
parton
structure function behaviour in the very small $x$ region whereas at
comparatively
small energies the results depend significantly on the values of the heavy
quark
mass and the QCD scale (see e.g. [4,5]).

Another approach to heavy flavour production processes has been considered in
Ref. [6] in the framework of the Quark--Gluon String Model (QGSM).
This model is a
version of the Dual Topological Unitarization (DTU)
and describes quite reasonably many features
of high energy production processes, including the inclusive spectra of
different secondary hadrons, their multiplicities, KNO--distributions, etc.,
both in hadron--nucleon and hadron--nucleus collisions [7--10].
High energy interactions are considered as proceeding via the exchange
of one or several pomerons and all elastic and inelastic processes
result from cutting pomerons or between pomerons [11].
The possibility of exchanging a different number of pomerons introduces
absorptive corrections to the inclusive cross sections which are in agreement
with the experimental data on production of hadrons consisting in light quarks.
The inclusive spectrum of heavy flavoured hadrons in QGSM was
considered earlier in Refs. [6,12,13].

In the present paper we discuss the possibility of using QGSM for
calculating heavy flavour production
at not very high energies. If
the ratio $m_h/\sqrt{s}$ is not negligibly small some corrections can
appear, which disappear with increasing energy. We
consider several variants of such corrections and compare the results with the
experimental data on charm and beauty production by proton and pion beams.
These corrections can also change the predicted $A$--dependence of charm and
beauty production. In particular we present quantitative
predictions for $B$--meson and $\Lambda_{b}$ production by protons on
nucleon and tungsten targets at HERA--B [14] energy,
which we assume to be equal
$\sqrt{s} =$ 39 GeV (i.e., $p_{lab} \approx $ 800 GeV/c).

We also present the results on beauty production at HERA--B energy obtained
with
a QGSM--based Monte Carlo code in which each inelastic collision is considered
as a hard
gluon--gluon one, simulated by PYTHIA. The $A$--dependence of beauty production
in hadron-nucleus collisions obtained in this case is very different from two
of the three variants of the low energy corrections considered in the
analytical calculations.

\vskip 0.9 truecm
\noindent{\bf 2. INCLUSIVE SPECTRA OF HEAVY FLAVOURED HADRONS IN QGSM}
\vskip 0.5 truecm

As mentioned above high energy hadron--nucleon and
hadron--nucleus interactions are considered in the QGSM as proceeding via
the exchange of one or several pomerons. Each pomeron corresponds to a
cylindrical diagram so in the case of cutting a pomeron two showers of
secondaries are produced, see Fig. 1. The inclusive spectrum of secondaries is
determined by the convolution of diquark, valence and sea quark distributions
$u(x,n)$ in the incident particles and the fragmentation functions $G(z)$
of quarks and
diquarks into secondary hadrons. The diquark and quark distribution
functions depend on the number $n$ of cut pomerons in the considered
diagram. In the case of a nucleon target the inclusive spectrum of a secondary
hadron $h$ has the form [7]:
\begin{equation}
1/\sigma_{inel} d\sigma/dx=\sum_{n=1}^{\infty}w_{n}\phi_{n}^{h}(x)\ \ ,
\end{equation}
where the functions $\phi_{n}^{h}(x)$ determine the contribution of diagrams
with $n$ cut pomerons and $w_{n}$ is the probability of this process.
Here we neglect the contributions of diffraction dissociation processes which
are comparatively small in the case of charmed and beauty hadron production.

For $pp$ collisions
\begin{equation}
\phi_{n}^{h}(x) = f_{qq}^{h}(x_{+},n)f_{q}^{h}(x_{-},n) +
f_{q}^{h}(x_{+},n)f_{qq}^{h}(x_{-},n) +
2(n-1)f_{s}^{h}(x_{+},n)f_{s}^{h}(x_{-},n)\ \  ,
\end{equation}
\begin{equation}
x_{\pm} = \frac{1}{2}[\sqrt{4m_{T}^{2}/s+x^{2}}\pm{x}]\ \ ,
\end{equation}
where $f_{qq}$, $f_{q}$ and $f_{s}$ correspond to the contributions of
diquarks, valence and sea quarks respectively.
They are determined by the convolution
of the diquark and quark distributions with the fragmentation functions, e.g.,
\begin{equation}
f_{q}^{h}(x_{+},n) = \int_{x_{+}}^{1} u_{q}(x_{1},n)G_{q}^{h}(x_{+}/x_{1})
dx_{1}\ \ .
\end{equation}
In the case of a meson beam the diquark contributions in Eq. (2) should be
changed by the contribution of valence antiquarks. The
diquark and quark distributions as well as the fragmentation functions are
determined via their Regge asymptotics with accounting for
conservation laws [7,9].

In present calculations we use quark and diquark distributions
in the proton of the form [7]:
\begin{eqnarray}
u_{uu}(x,n) &=& C_{uu}x^{2.5}(1-x)^{n-1.5}\ \ ,\\
u_{ud}(x,n) &=& C_{ud}x^{1.5}(1-x)^{n-1.5}\ \ , \\
u_{u}(x,n) &=& C_{u}x^{-0.5}(1-x)^{n+0.5}\ \ , \\
u_{d}(x,n) &=& C_{d}x^{-0.5}(1-x)^{n+1.5}\ \ ,\\
u_{\overline{u}}(x,n) &=& u_{\overline{d}}(x,n) = C_{\overline{u}}x^{-0.5}
\nonumber
[(1+\delta /2)\\&\times &(1-x)^{n+0.5}(1-x/3)-
\delta /2(1-x)^{n+1}] \; , \; n>1\ \ ,\\
u_{s}(x,n) &=& C_{s}x^{-0.5}(1-x)^{n+1} \; , \; n>1\ \ ,
\end{eqnarray}
where $\delta =0.2$ is the relative probability to find a strange quark in the
sea. The factors $C_{i}$ are determined from the normalization condition
\begin{equation}
\int_{0}^{1} u_{i}(x,n)dx = 1\ \ .
\end{equation}

The fragmentation functions of quarks and diquarks into charmed mesons and
baryons were changed a little in comparison with Ref. [12] to obtain a better
agreement with the existing experimental data. We use these functions in the
form
\begin{equation}
G_{u}^{\overline{D^{0}}} = G_{d}^{D^{-}} =
a_{0}(1-z)^{\lambda-\alpha_{\psi}(0)}(1+a_{1}z^{2})\ \ ,
\end{equation}
\begin{equation}
G_{u}^{D^{-}} = G_{u}^{D^{+}} = G_{u}^{D^{0}} = G_{d}^{D^{+}} =
G_{d}^{D^{0}} = G_{d}^{\overline{D^{0}}} =
a_{0}(1-z)^{1+\lambda-\alpha_{\psi}(0)}\ \ ,
\end{equation}
\begin{equation}
G_{uu}^{D^{+}} = G_{uu}^{D^{-}} = G_{uu}^{D^{0}} =
G_{ud}^{D^{+}} = G_{ud}^{D^{0}} =
a_{0}(1-z)^{3+\lambda-\alpha_{\psi}(0)}\ \ ,
\end{equation}
\begin{equation}
G_{uu}^{\overline{D^{0}}} =
a_{0}(1-z)^{2+\lambda-\alpha_{\psi}(0)}(1+a_{2}z^{2})\ \ ,
\end{equation}
\begin{equation}
G_{ud}^{\overline{D^{0}}} =
a_{0}(1-z)^{2+\lambda-\alpha_{\psi}(0)}(1-z+a_{2}z^{2}/2)\ \ ,
\end{equation}
\begin{equation}
G_{uu}^{\Lambda_{c}} = G_{ud}^{\Lambda_{c}} =
a_{01}(1-z)^{6+\lambda-\alpha_{\psi}(0)}\ \ ,
\end{equation}
\begin{equation}
G_{u}^{\Lambda_{c}} = G_{d}^{\Lambda_{c}} =
a_{01}(1-z)^{2+\lambda-\alpha_{\psi}(0)}\ \ ,
\end{equation}
\begin{equation}
G_{\overline{u}}^{\Lambda_{c}} = G_{\overline{d}}^{\Lambda_{c}} =
G_{u}^{\Lambda_{c}}(1-z)\ \ .
\end{equation}

In the case of $\Lambda_{c}$ production there are two different contributions
[8]. The first one corresponds to the central production of a
$\Lambda_{c}\overline{\Lambda}_{c}$ pair and can be described by the
previous formulas. The second contribution is connected with the direct
fragmentation of the initial baryon into $\Lambda_{c}$ with conservation
of the string junction. To account for this possibility we input into Eq. (2),
instead of $f_{q}(x_{+},n)$ and
$f_{q}(x_{-},n)$,
two additional items $f_{qq2}(x_{+},n)$ and $f_{qq2}(x_{-},n)$
multiplied by $(1-x_-)^{-\alpha_\psi (0)}$ and $(1-x_+)^{-\alpha_\psi (0)}$
respectively.
The values of $f_{qq2}(x_{+},n)$ and $f_{qq2}(x_{-},n)$ are
determined by the corresponding fragmentation functions
\begin{equation}
G_{uu2}^{\Lambda_{c}} = a_{02}z^{2}(1-z)^{1+\lambda-\alpha_{\psi}(0)}\ \ ,
\end{equation}
\begin{equation}
G_{ud2}^{\Lambda_{c}} = a_{02}z^{2}(1-z)^{\lambda-\alpha_{\psi}(0)}\ \ .
\end{equation}

The probability for a process to have $n$ cutted pomerons was calculated
using the quasi\-ei\-ko\-nal approximation [7,15]:
\begin{equation}
w_{n} = \sigma_{n}/\sum_{n=1}^{\infty}\sigma_{n}  \;  , \;
\sigma_{n} = \frac{\sigma_{P}}{nz} (1-e^{-z} \sum_{k=0}^{n-1}\frac{z^{k}}{k!
})\ \  ,
\end{equation}
\begin{equation}
z = \frac{2C\gamma}{R^{2}+\alpha^{\prime}\xi}e^{\Delta\xi} \; ,
\; \sigma_{P} = 8\pi
\gamma e^{\Delta\xi} \; , \; \xi = \ln(s/1\ {\rm GeV}^{2})  \ \ ,
\end{equation}
with parameters
\vskip 0.3 truecm
\begin{center}
$\Delta = 0.139 \; , \; \alpha^{\prime} = 0.21\ {\rm GeV}^{-2} \; ,
\; \gamma_{pp} = 1.77
\; , \; \gamma_{\pi p} = 1.07 \; ,$\newline
$R_{pp}^{2} = 3.18\ {\rm GeV}^{-2} \; , \; R_{\pi p}^{2} = 2.48\ {\rm GeV}
^{-2} \;
, \; C_{pp} = 1.5 \; , \; C_{\pi p} = 1.65\ \ .$
\end{center}
\vskip 0.3 truecm

In the case of secondary production at not very high energy or if the mass
of the secondary hadron is comparatively large some problems appear
in QGSM. Their
origin is connected with the normalization (to unity) of quark and
diquark distributions.
The probability
for any quark or diquark to fragment into a secondary hadron $h$ is also
normalized to unity,
\begin{equation}
\sum_{h} \int^{1}_{0} G^h_i(z) G^h_i(0) dz = 1\ \ .
\end{equation}
However the lower limit of integration in Eq. (4) is higher than zero. The
minimal value of $x_1$ is $x_{1 min} \approx m^2_T/s$, so the quark having
$x_1 < x_{1 min}$ cannot fragment into a secondary hadron. As a result the
sum of the energy of all secondaries becomes slightly smaller than the initial
energy. The problem is more clear in the case of a quark which exists in the
initial hadron. For example in the case of $Dp$ collisions at not very high
energy the multiplicity of secondary charmed hadrons will be
smaller than unity. QGSM is based on the Regge theory which contains
many other corrections of the same order $m^2_h/s$ which have been omitted.
However even numerically a small nonconservation of energy and quantum
numbers can produce problems. To avoid them we consider two possibilities to
correct QGSM. The simplest way is to use the normalization condition
\begin{equation}
\int_{m^2_T/s}^{1} u_{i}(x,n)dx = 1
\end{equation}
instead of Eq. (11). Another possibility is to multiply every $f_i$ in
Eq. (2) by a factor
\begin{equation}
D_n^{+} = \int^{1}_{0} f_{i}^{h}(x,n) dx
\Bigm / \int^{1}_{0} f_{i}^{h}(x_+,n) dx
\end{equation}
and
\begin{equation}
D_n^{-} = \int^{0}_{-1} f_{i}^{h}(x,n) dx
\Bigm / \int^{0}_{-1} f_{i}^{h}(x_-,n) dx\ \ .
\end{equation}

\vskip 0.9 truecm
\noindent{\bf 3. CHARMED AND BEAUTY HADRON PRODUCTION ON NUCLEON
TARGET WITH ACCOUNT FOR
LOW ENERGY CORRECTIONS}
\vskip 0.5 truecm

It is well known that perturbative QCD taking into account leading order
$(\sim \alpha_s^2)$ and next--to--leading order
$(\sim \alpha_s^2)$ contributions
describes quit well the total cross section of heavy flavour production
on nucleon
targets. However there is a serious problem to reproduce simultaneously the
experimental cross sections of $D$--mesons [16] and $\Lambda_c$'s [17]
as well as the shapes of their $x_F$--distributions. In particular,
the Parton Fusion Model [18] assumes that the produced charmed quarks
fragment into hadrons thus transferring them some fraction of their momentum
and
cannot recombine with valence quarks because of the large rapidity gap between
charmed and valence quarks. This model predicts the same $x_F$--distribution
for
favoured and unfavoured $D$--mesons in disagreement with data [16,19]. Taking
into
account the intrinsic charm contribution [20] allows one [21] to describe the
$x_F$--spectra of $D$--mesons as well as the shape of the $\Lambda_c$ spectrum.
However the $\Lambda_c$ spectrum at moderate $x_F$ becomes more than two
orders of magnitude smaller than the experimental data [17].

QGSM does not have so large problems to describe the data [16,17] on $D$ and
$\Lambda_c$ production [12]. The largest difference is
not greater than 2--3 times, which is not very bad
keeping in mind the normalization
uncertainties. The results of three sets of calculations, without
any corrections
(QGSMa), with the corrections (25) (QGSMb) and with the corrections (26),
(27) (QGSMc),
are compared with the experimental data
on the cross sections for total charm
and for different charmed hadrons in Table 1. The model
parameters for these sets are presented in Table 2. Let us note that the value
of $\alpha_\psi(0)$ is comparatively well known [6,12,22] whereas the value of
$\alpha_{\Upsilon}(0)$ can be changed significantly. For simplicity we use the
same parameter values in the cases QGSMa and QGSMb, where the numerical
difference of the results is not large.

Let us compare now the results of our calculations with experimental data
(Table 1). One can see that the energy
dependence of charm production cross section in QGSMb and especially in QGSMc
cases is not so strong as in QGSMa. Let us note also that
the QGSMc energy dependence is significantly weaker than perturbative QCD
predictions if one use modern (i.e., singular at small $x$) gluon distribution
and the same scale values at different energies. In all variants we can see a
reasonable agreement with the total cross section of charm production in $pp$
collisions at 400 GeV/c [16] and 800 GeV/c [23]. In the case of
$Sp\overline{p}S$
energy [24], the experimental errors are too large for excluding any variants.
In the cases of QGSMa and QGSMb the calculated cross section for $D^{0}$ and
$D^{+}$ production seems to be too small whereas the cross section of
$\overline{D^{0}}$ production is too large. The reasons for this large
difference in the
calculated values of $\sigma(D)$ and $\sigma(\overline{D})$ are the large
values of the parameters $a_{1}$ and $a_{2}$ in Eq.(8) which is connected
with our
wish to obtain a comparatively large value of $\sigma(\Lambda_{c})$ at
$\sqrt{s}$ = 62 GeV. Correspondingly the cross section of $\Lambda_{c}$
production at 400 GeV/c [16] is also too large. In the case of QGSMc the
parameter $a_{0}$ can not be taken significantly smaller that the used value
to avoid too small a cross section at $\sqrt{s}$ = 630 GeV. So the values of
$a_{1}$ and $a_{2}$ can not be taken so large as in the other cases.
As a result
the $D^{0}$ and $D^{+}$ cross sections are larger whereas
$\sigma(\Lambda_{c})$ and $\sigma(\overline{D})$ are closer to the
experimental data. Let us remark also the large difference between old [23]
and new [25,26] data on $D$--meson production at 800 GeV/c.

The shapes of the inclusive spectra of $D$--mesons do not differ
significantly with those of [12] as one can see in Fig. 2.

In the case of beauty hadron production we do not have enough data to fix the
parameters of the fragmentation functions. So we use the same
fragmentation functions as for charm production with
parameters $b_{i}$ (which are also presented in Table 2) instead of
$a_{i}$ in Eqs. (12)--(21). Namely for $B^{0}$ production we use the same
fragmentation function as for $D^{+}$ , for $B^{-}$ as for $D^{0}$ , for
$\overline B^{0}$ as for $D^{-}$, for $B^{+}$ as for $\overline D^{0}$ and for
$\Lambda_{b}$ as for $\Lambda_{c}$.

The existing experimental point on the total beauty production cross section at
$\sqrt{s}$ = 630 GeV [27] allow us to fix the parameter $b_{0}$. The values of
$b_{1}$ and $b_{2}$ as well as $b_{01}$ and $b_{02}$ can be estimated from
the condition that the cross section of $\Lambda_{c}$ production should be
(at not very high energy) slightly smaller than the difference of antibeauty
and
beauty production cross section (including the case of a nuclear target).
Naturally
the low energy corrections are more important in the case of beauty production
and the difference in the predicted energy dependence between the three
variants is larger, as one can see in Table 3. Actually we normalize
the beauty production cross section at $\sqrt{s}$ = 630 GeV using the
experimental
point [27] and obtain very different predictions for
low energy $\pi p$ collisions. The variants QGSMa and QGSMb
predict the last cross section to be very small compared to the data
[28, 29] and the variant QGSMc predicts it to be slightly too large. In the
case of beauty production in $pp$ collisions at 800 GeV/c [30] the data are in
agreement with QGSMa and QGSMb and in strong disagreement with QGSMc. However
in Refs. [28--30] nuclear targets were used
and the values presented in Table 3 for nucleon targets are the results of a
linear extrapolation. So the experimental values possibly can be increased if
the $A$--dependence of beauty production is weaker than $A^1$.

Our predictions for beauty production in $pp$ interactions at
$\sqrt{s}$ = 39 GeV (HERA--B energy) are presented in Table 4. They are in
reasonable agreement with the results of the similar calculations of Ref. [13],
where the cross section for beauty production
was equal to 0.01--0.1 $\mu$b depending on the
parameters.

\vskip 0.9 truecm
\noindent{\bf 4. CHARM AND BEAUTY PRODUCTION ON NUCLEAR TARGETS}
\vskip 0.5 truecm

The models based on the DTU approach allow one to calculate the inclusive
spectra
of secondaries produced on nuclear targets [31--34] without new
phenomenological parameters. In multiple scattering theory a
hadron--nucleus interaction can be considered as a superposition of
interactions
with a different number of separated nucleons. In the QGSM we account for the
possibility of multipomeron interaction in every NN blob [10,34]. One cut
pomeron connects a valence quark--diquark pair of the beam nucleon with a
quark--diquark
pair of a target nucleon. All other pomerons connect sea quark--antiquark
pairs of the incident nucleon with valence quark--diquark pairs or sea
quark--antiquark pairs of the target nucleons, see Fig. 3. To account
for that
in our calculation of quark and diquark distributions in the beam nucleon the
value of
$n$ in Eqs. (5)--(10) is equal to the total number of cutted pomerons in the
nucleon--nucleus interaction, and in the same distributions for a target
nucleon
the corresponding value of $n$ is equal to the number of pomerons connected
with a given nucleon. It provides the asymmetry of the inclusive spectra of
secondaries produced on nuclear targets in comparison with a nucleon--nucleon
interaction. We assume that the secondaries are produced ndependently in
every shower
and their formation zone is large in comparison with the nuclear
radius. The contribution of intranuclear cascade to the spectra of
secondaries, as well as the Fermi motion of target nucleons, is neglected.

A detailed description of light flavour hadron yields from nuclear targets in
the QGSM was presented in Refs. [10,34]. Now we use the same probability of
interaction with a given number of target nucleons and consider only the cross
sections and the spectra of beauty hadrons.

In the case of QGSMa the charm and beauty cross section
in $pA$ collisions has a $ A^{\alpha } $--dependence with very small
$\alpha$ [10]. It is a consequence of the very strong energy dependence
of heavy
flavour production cross section at comparatively small energies. The
initial energy is divided between several cut pomerons so the effective
energy of every shower decreases and the corresponding contribution to charm
and beauty production cross section is small. So really only the interaction
with one target nucleon gives the main contribution to the production cross
section in QGSMa. In the case of QGSMb we have approximately the same
behaviour because the corrections are comparatively small. However in the case
of corrections (26), (27) the situation changes and in this case (QGSMc) the
value of $\alpha $ becomes closer to unity (see Table 5,
where we present the calculated values of $\alpha $ for the total charm and
beauty production cross section in the interval $A = 1 \div 208 $).

The predictions of QGSMc for the inclusive spectra of $B$--mesons and
$\Lambda_{b}$'s produced in $pp$ and $pW$ collisions at $\sqrt{s}$ = 39 GeV
are presented in Fig. 4. As expected, the cross section
of all $B$--mesons
decrease with $x_F$ faster than that of
$\Lambda_{b}$'s. The yields from
a nucleus target are larger at small $x_F$ and smaller at large $x_F$, as
usual.
Some difference in the nuclear effects for $B$--mesons and $\Lambda_{b}$'s can
be explained by the diffusion of the last ones from the nuclear fragmentation
region to the central region as a consequence of the comparatively
small energy.

\vskip 0.9 truecm
\noindent{\bf 5. MONTE CARLO RESULTS}
\vskip 0.5 truecm

In this Section we are going to present the results on beauty production
obtained with a QGSM--based Monte Carlo code (a detailed description can be
found in Ref. \cite{MCNF}). This Monte Carlo model describes hadron--hadron,
hadron--nucleus and nucleus--nucleus collisions on the same footing, at the
partonic level.
The nuclear parton wave function is constructed as a convolution of parton
distributions of individual nucleons with the distribution of nucleons inside
the nucleus. The position of each nucleon is taken to be described by the
Woods--Saxon density:
\begin{equation}
\rho (r)\ =\rho _0/(1+ \exp [(r - r_0)/a])\ \ ,
\label{5-1}
\end{equation}
with
\begin{equation}
r _0 = 1.19 A^{1/3}+ 1.61 A^{-1/3}\ {\rm fm},\mbox{  } a\mbox{ } =\mbox{ }
0.54\ {\rm fm}.\label{5-2}
\end{equation}

To take the Fermi motion of nucleons into account we generate a Fermi momentum
$p$ for each nucleon uniformly distributed in the range $0 < p
< p_{F}$, where
$p_{F}$ is the maximum Fermi nucleon momentum:
\begin{equation}
p_F=(3{\pi }^2)^{1/3}h{\rho }^{1/3}(r)\ \ ,
\label{5-3}
\end{equation}
with $h = 0.197$ fm$\ $GeV/c. Isotropical angular distribution is assumed in
both the coordinate and momentum spaces.

As to the parton distribution for individual nucleons this
is taken to be the same as for $NN$ collisions. In particular, the
distribution in the number of partons in a nucleon,
which
is directly connected with the value of the multipomeron vertices in the
reggeon
theory (\cite{AGGK}), is taken poissonian:
\begin{equation}
\label{5-4}w_{N} = \exp(-g(s) )g ^N (s) / {N!}\ \ ,
\end{equation}
corresponding to the eikonal picture. The mean number of partons
in each nucleon,
$g(s) = g_0 s^{\Delta }$, is a function of the center of mass energy
$\sqrt{s}$.
We use
$g_{0} = 3.0$ and $\Delta = 0.09$.

The parton distribution in
impact parameter (relative to the center of the corresponding nucleon) is
taken to be gaussian, in accordance with the Pomeron picture of strong
interactions:
\begin{equation}
\label{5-5}F(b_p)=(4\pi \lambda )^{-1}\exp (-b_p^2/4\lambda )\ \ ,
\end{equation}
with the radius depending on the initial nucleon energy. For a projectile
or target nucleon, $\lambda = R^{2} + \alpha ^{\prime }\ln{\sqrt{s}}$, where
$\alpha ^{\prime } = 0.01\ {\rm fm}^{2}$ and $R^{2} = 0.15 \ {\rm fm}^{2}$.

A hadron or nucleus collision is assumed to be the
interaction between partons from the projectile and target. A parton from the
projectile can interact with one from the target if they lie in impact
parameter space within an area determined by the parton--parton cross section,
which has been assumed energy independent,
$\sigma _p = 3.5$ mb.

In this way the number of inelastic collisions is
determined and
the inelastic cross section is calculated (an elastic event is one with no
partons close enough in impact parameter space).

Now, opposite to Ref. \cite{MCNF} (where each inelastic collision could be a
hard or a soft one with a probability $w(s)$), each inelastic collision
is taken to be a hard gluon--gluon one. For simulating
these $gg$ collisions the PYTHIA program (\cite{PYTHIA}) is used. Only
$gg\longrightarrow gg$, $gg\longrightarrow q\overline{q}$ and
$gg\longrightarrow Q\overline{Q}$ collisions, with $q$
and $Q$ light and heavy flavours
respectively, are considered. In our calculations we have used the EHLQ
set 1 structure
functions (\cite{EHLQ84}), a $K=2.0$ factor and a $p_{t}^{min} = 2.3$ GeV/c.

Using this model we have computed the $x_F$-- and $p_t$--distributions of
particles with beauty in $pp$ and $pW$ collisions at $\sqrt{s}=39$ GeV
(Figs. 5 and 6). The number of generated events is 600000 for $pp$ and 400000
for $pW$ collisions. Also in the last column of Table 5 we present the Monte
Carlo results for the $A$--dependence of charm and beauty cross sections.

\vskip 0.9 truecm
\noindent{\bf 6. CONCLUSIONS}
\vskip 0.5 truecm

In this paper we have presented results of QGSM for heavy flavour production at
different energies. Three variants of QGSM have been described,
which try to take
into account corrections which are important at not very high energies.
Besides, results of a Monte Carlo code \cite{MCNF}
for production of beauty particles at HERA--B energy have been shown.

We can see that the three variants of QGSM describe reasonably most of
the existing data on charmed hadron production. Some contradictions are
connected with the significant differences of different sets of experiment
data.
For example, the cross section of $\Lambda_{c}$ production measured in
Ref. [16] is very small in comparison with the data of Ref. [17]. If the
$\Lambda_{c}$ production cross section is really so small the difference in
$\overline{D}$ and $D$ production cross sections could be obtained smaller by
decreasing the parameters $a_{1}$ and $a_{2}$ in Eqs. (13), (16) and (17).

In the case of beauty production at $\sqrt{s}$ = 39 GeV the model cannot
describe simultaneously the data obtained with proton and pion beams [28--30].
As
one can see in these experimental papers, perturbative QCD also can not
describe
them with the same parameters (QCD scale and quark masses). In our case we can
describe the proton beam data in the cases of QGSMa and QGSMb and the pion
beam data are in reasonable agreement with QGSMc. A possible explanation of
such situation can be the assumed $A$--dependence of beauty production, which
could be
weaker than $A^1$. However this would be in disagreement with the direct
measurement
of the $A$--dependence of $D$--meson production [26] where
$\alpha = 1.02 \pm 0.03 \pm 0.02$.

So the variants QGSMa and QGSMb predict the beauty production cross section
in $pp$ collisions at $\sqrt{s}$ = 39 GeV equal to $5 \div 7$ nb in agreement
with data [30] and in this case they cannot describe the data of beauty
production by pion beams [28, 29]. The variant QGSMc is in reasonable agreement
with the data [29], gives slightly too large cross sections in comparison with
the data [28] and predicts the cross section of beauty production in $pp$
collisions at $\sqrt{s}$ = 39 GeV equal to 0.2--0.3 $\mu$b with
an $A$--dependence
$\sim A^{0.99}$. The $A$--dependence obtained in the Monte Carlo calculations
is
$A^{0.89}$, closer the the variant QGSMc than to QGSMa or QGSMb.
\vskip 0.3 truecm

We are grateful to R. Rueckl for stimulating discussions. We also thank the
Xunta de Galicia, the Direcci\'on General de Pol\'{\i}tica Cient\'{\i}fica and
the CICYT of Spain for financial support.

\pagebreak

\newpage

\begin{center}
{\bf Table 1}\vspace{15pt}
\end{center}
Comparison of the experimental cross sections (in $\mu$b) of different charmed
hadrons produced in $pp$ and $\pi p$ interactions with the results of our
QGSM calculations. In the case of $\pi p$ collisions, the cross
sections are given only in the region $x_{F} > 0$ (i.e., in the outgoing $\pi $
hemisphere).
\vskip 12pt
\begin{center}
\begin{tabular}{|c||c|c|c|c|}\hline

Reaction, $\sqrt{s}$  & Experiment  &QGSMa  &QGSMb  &QGSMc  \\   \hline

$\pi^{-} p \rightarrow D^{+}/D^{-}$, 26 GeV
&5.7$\pm$1.6   &9.7    &10.7  & 11.2     \\

$\pi^{-} p \rightarrow D^{0}/\overline D^{0}$, 26 GeV
&10.1$\pm$2.2   &6.3   &7.1  & 10.3   \\

$pp \rightarrow c \overline c$, 27 GeV  & 14--23
&28   &33   & 30   \\

$pp \rightarrow D^{+}/D^{-}$, 27 GeV  &11.9$\pm$1.5  & 12
 &14.2   &23.9  \\

$pp \rightarrow D^{0}/\overline D^{0}$, 27 GeV  &18.3$\pm$2.5
&23.4   &27.4   &26.6   \\

$pp \rightarrow D^{0}$, 27 GeV   &10.5$\pm$1.9
&3.7   &4.4   &10.5   \\

$pp \rightarrow \overline D^{0}$, 27 GeV   &7.9$\pm$1.5
&19.7   &23   &16.1   \\

$pp \rightarrow D^{+}$, 27 GeV   &5.7$\pm$1.1
&3.7  &4.4   &10.5   \\

$pp \rightarrow D^{-}$, 27 GeV    &6.2$\pm$1.1
&8.3   &9.8   &13.4    \\

$pp \rightarrow \Lambda_{c}$, 27 GeV   &$< 6.1\;(\Lambda_{c}/D)$
&14.6   &14.6   &7.3   \\

--   &$< 15\;(\Lambda_{c}/\overline \Lambda_{c})$
 & -- & -- & --   \\

$pp \rightarrow c \overline c$, 39 GeV   & 29--55
&47   &52   &41   \\

$pp \rightarrow D^{+}/D^{-}$, 39 GeV  &33$\pm$7   &25
&27   &34   \\

$pp \rightarrow D^{0}/\overline D^{0}$, 39 GeV   &$26^{+21}_{-13}$
&40   &44   &38   \\

$pp \rightarrow D^{0}/\overline D^{0}$, 39 GeV   &17.7$\pm$0.9$\pm$3.4
& --  & --  & --  \\

$pp \rightarrow D^{0}$, 39 GeV   &38$\pm$3$\pm$13
& 6.5 & 10 & 15.5 \\

$pp \rightarrow D^{+}$, 39 GeV   &38$\pm$9$\pm$14
& 6.5  & 10 & 15.5 \\

$pp \rightarrow \Lambda_{c}$, 62 GeV  &(40$\pm 18) \div (204\pm$91)
 &19   &19   &8.9     \\

$pp \rightarrow c\overline c$, 630 GeV & 680$\pm 560 \pm 250 \pm$210
&660   &660   &415   \\   \hline
\end{tabular}
\end{center}

\newpage

\begin{center}
{\bf Table 2}
\end{center}
\vspace{15pt}
The values of the parameters used for the calculations of charm and beauty
production in QGSM.
\begin{center}
\vskip 12pt
\begin{tabular}{|c||c|c|c|}\hline

 Parameter   &QGSMa   &QGSMb  &QGSMc   \\   \hline

$\alpha_\psi(0)$  & $-2$  & $-2$  & $-2$    \\

$a_0$    & 0.024   & 0.024  & 0.02   \\

$a_{1}$  & 10  & 10   & 0  \\

$a_{2}$  & 50  & 50  & 16   \\

$a_{01}$  & 0.011  & 0.011  & 0.007   \\

$a_{02}$  & 0.005  & .005  & 0.0025   \\

$\alpha_{\Upsilon}(0)$  & $-8$  & $-8$  & $-8$  \\

$b_0$    & 0.011   & 0.011  & 0.0055   \\

$b_{1}$  & 5  & 5   & 6  \\

$b_{2}$  & 25  & 25  & 40   \\

$b_{01}$  & 0.005  & 0.005  & 0.0015   \\

$b_{02}$  & 0.0004  & .0004  & 0.0018  \\  \hline
\end{tabular}
\end{center}

\vskip 3.0cm
\begin{center}
{\bf Table 3}
\end{center}
\vspace{15pt}
Comparison of the experimental cross sections of beauty production
in $pp$ and $\pi p$ interactions with the results of our present QGSM
calculations.
\begin{center}
\vskip 12pt
\begin{tabular}{|c||c|c|c|c|}\hline

Reaction, $\sqrt{s}$  & Experiment   &QGSMa    &QGSMb  &QGSMc   \\   \hline

$\pi^{-} p \rightarrow b\overline{b}$, 31 GeV
&75$\pm$31$\pm$26 nb   & 2.4 nb  & 2.8 nb  & 125 nb     \\

$\pi^{-} p \rightarrow b\overline{b}$, 31 GeV , $x_F >0$
&47$\pm$19$\pm$14 nb   & 0.8 nb  & 1.0 nb  & 63 nb     \\

$\pi^{-} p \rightarrow b\overline{b}$, 34 GeV
&33$\pm$11$\pm$6 nb   & 4.7 nb  & 5.5 nb  & 154 nb     \\

$pp \rightarrow b\overline{b}$, 39 GeV & 5.7$\pm 1.5 \pm 1.3$ nb
& 4.8 nb  & 6.8 nb  & 270 nb    \\

$pp \rightarrow b\overline{b}$, 630 GeV & 19.3$\pm 7 \pm 9\ \ \mu$b
& 21 $\mu$b  & 21 $\mu$b  & 8 $\mu$b    \\   \hline
\end{tabular}
\end{center}

\pagebreak

\begin{center}
{\bf Table 4}
\end{center}
\vspace{15pt}
Predictions for beauty production in $pp$ interactions at $\sqrt{s}$ =
39 GeV. All the cross sections are in nb.
\begin{center}
\vskip 12pt
\begin{tabular}{|c||c|c|c|}\hline

Reaction   &QGSMa   &QGSMb  &QGSMc   \\   \hline

$pp \rightarrow b\overline{b}$  & 4.8 & 6.8 & 270  \\

$pp \rightarrow B^{+}/B^{-}$  & 3.9 & 5.5 & 220  \\

$pp \rightarrow B^{0}/\overline{B^{0}}$ & 2.1 & 3 & 210  \\

$pp \rightarrow \Lambda_{b}$  & 1.1 & 1.1 & 80  \\   \hline
\end{tabular}
\end{center}

\vskip 3.0cm
\begin{center}
{\bf Table 5}
\end{center}
\vspace{15pt}
Predictions for the $A$--dependence ($\sigma \sim A^\alpha$) of charm and
beauty
production on nuclear targets in the interval $A = 1 \div 208$ at
$\sqrt{s}$ = 39 GeV.
\begin{center}
\vskip 12pt
\begin{tabular}{|c||c|c|c|c|}\hline

Parameter   &QGSMa   &QGSMb  &QGSMc & Monte Carlo  \\  \hline
$\alpha(c\overline{c})$  & 0.74 & 0.75 & 0.85  & 0.87\\
$\alpha(b\overline{b})$  & 0.56 & 0.57 & 0.99  & 0.89\\    \hline
\end{tabular}
\end{center}

\pagebreak

\begin{center}
{\bf Figure captions}
\end{center}
\vskip 0.5 truecm

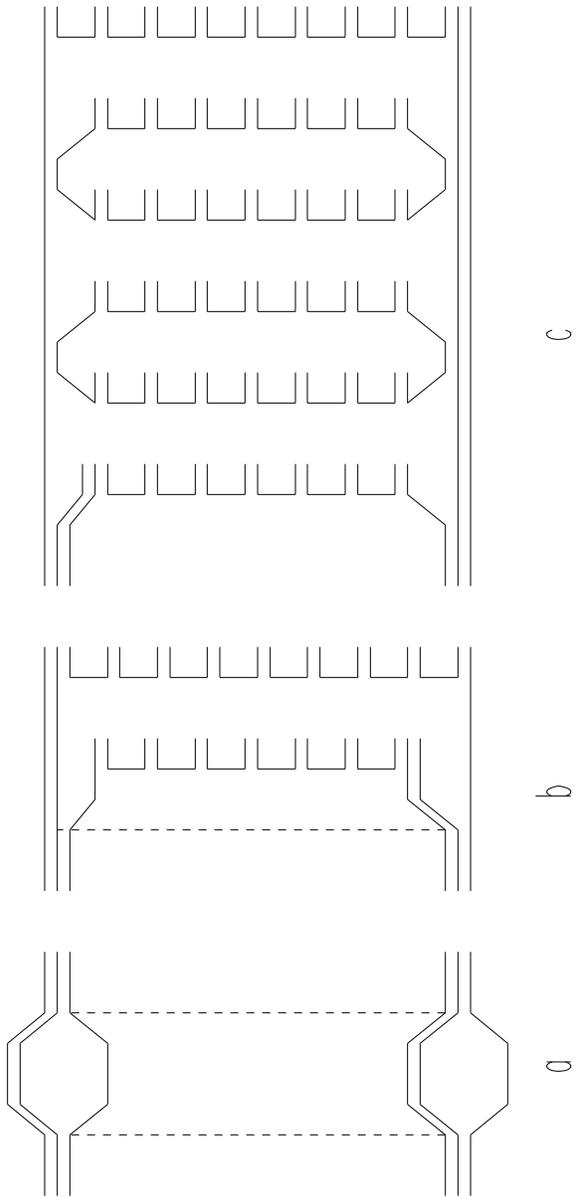
{\bf Fig.1.} Cylindrical diagram which corresponds to the one--pomeron exchange
contribution to elastic pp scattering (a). Its cut which determine the
contribution to inelastic pp cross section (b). The diagram which
correspond to the cut of three pomerons (c).

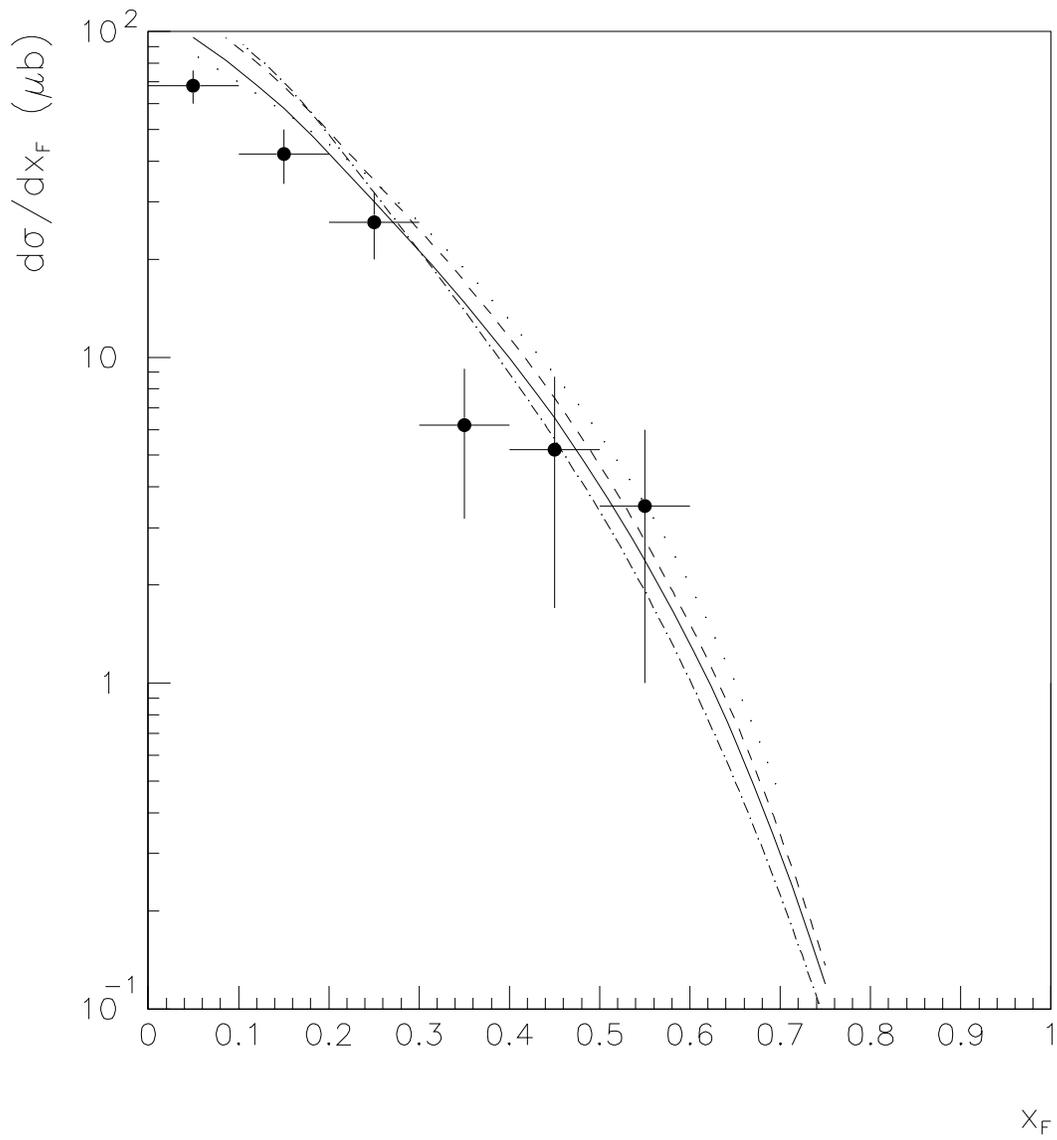
{\bf Fig.2.} Inclusive spectra of all $D$--mesons produced in $pp$ interactions
at 400 GeV/c [16]  and its description by QGSMa (solid curve), QGSMb (dashed
curve) and QGSMc (dashed--dotted curve). The dotted curve represent the
result of Ref. [12].

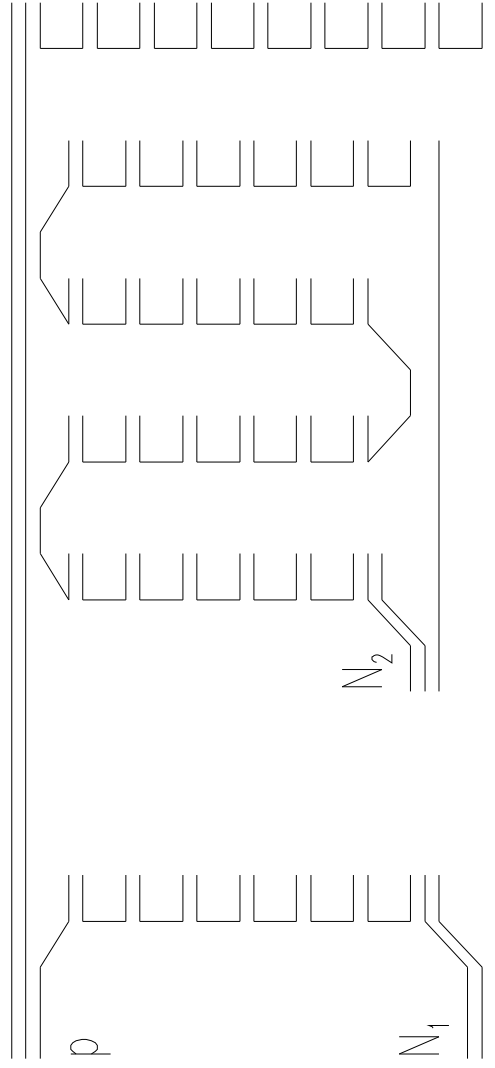
{\bf Fig.3.} Example of a diagram of interaction of the beam nucleon with two
target nucleons.

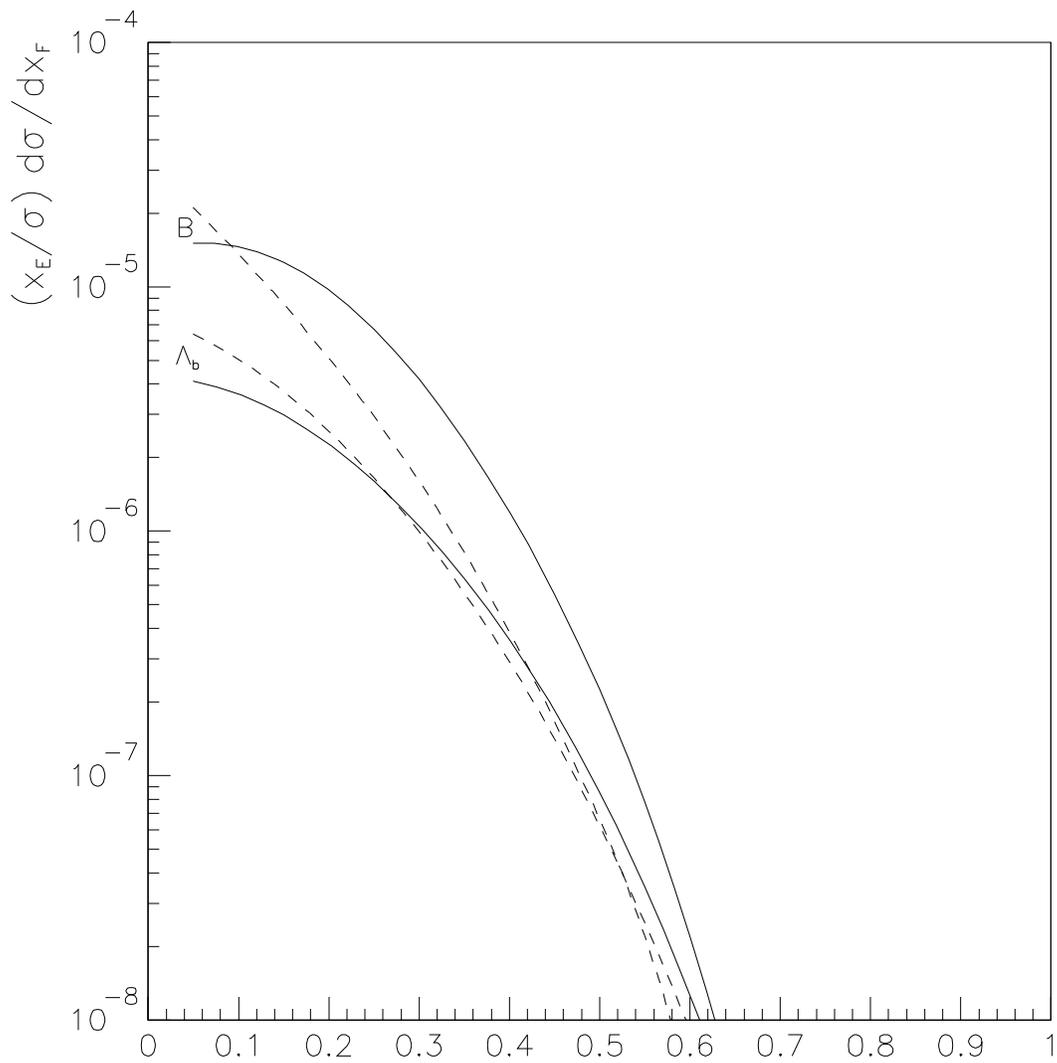
{\bf Fig.4.} Spectra of all $B$--mesons
and $\Lambda_{b}$'s
at $\sqrt{s}$ = 39 GeV for a proton beam on hydrogen (solid curves) and lead
(dashed curves) targets using the QGSMc.

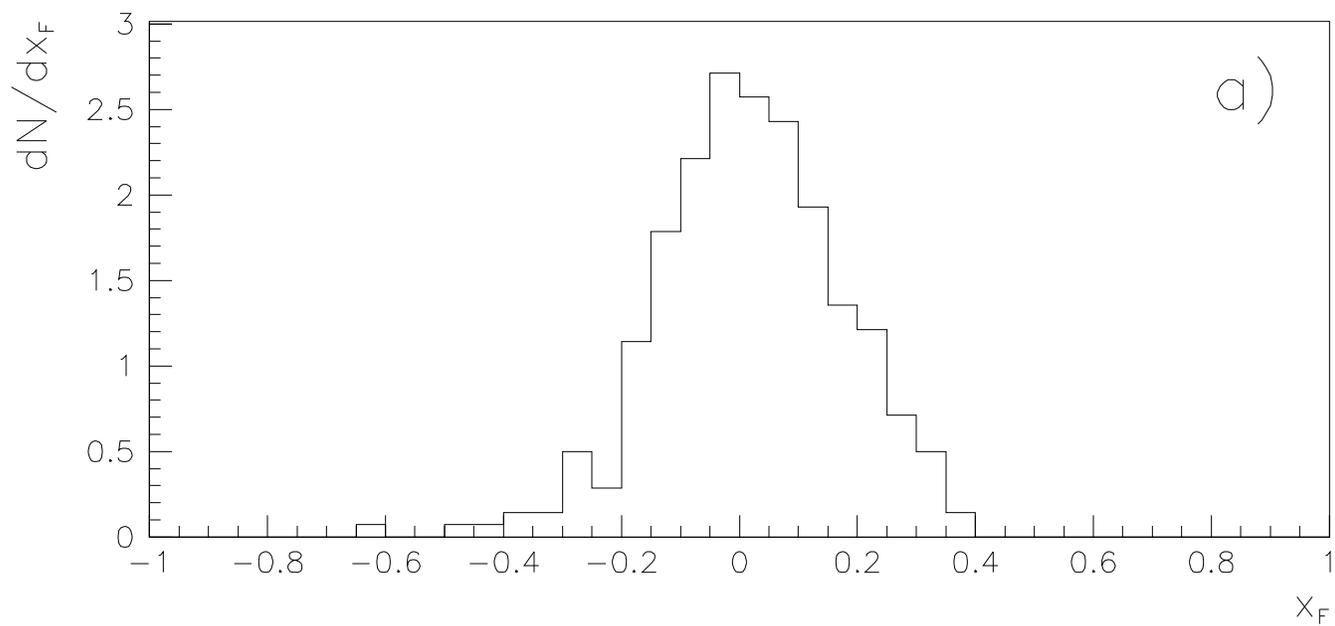
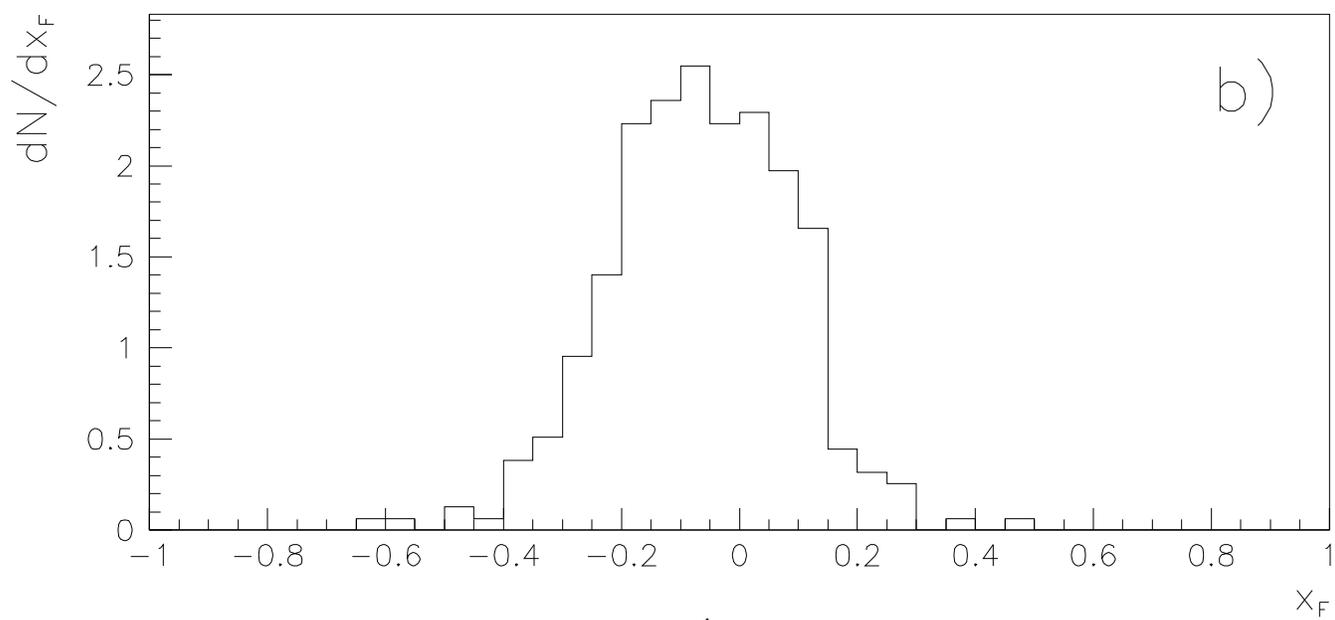
{\bf Fig. 5.} $x_F$--distributions (normalized to unity)  of beauty particles
in: a) $pp$ and b) $pW$
collisions at $\sqrt{s}=39$ GeV.

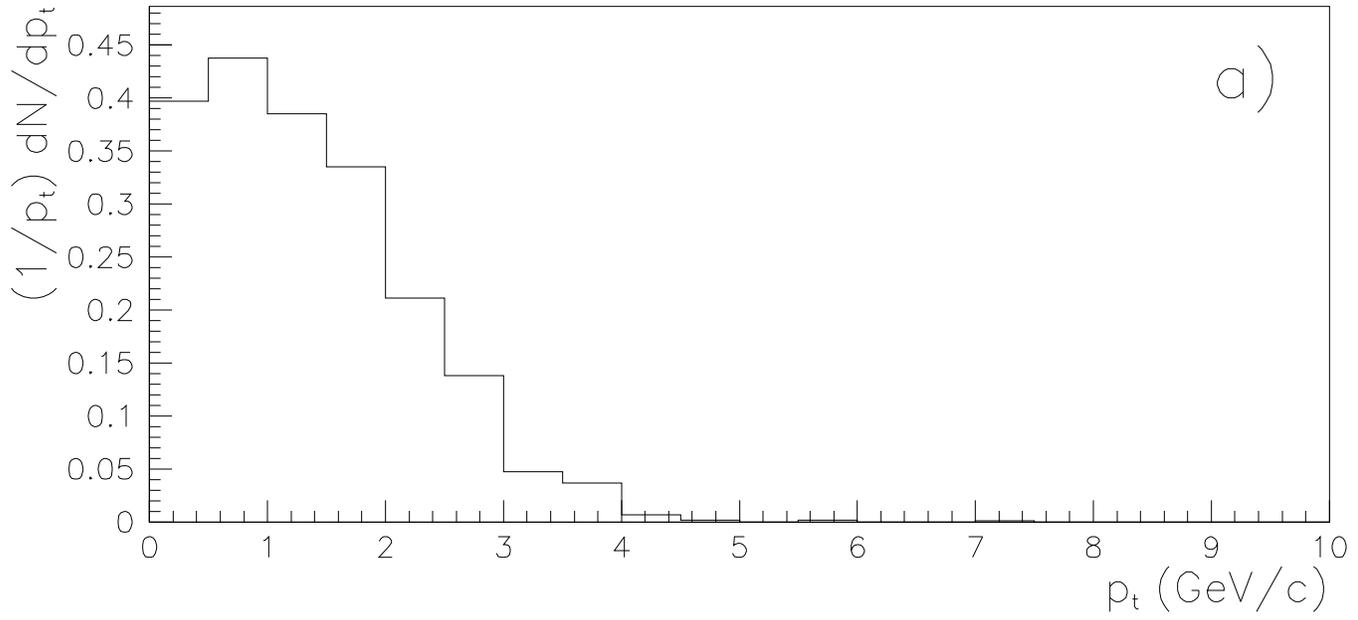
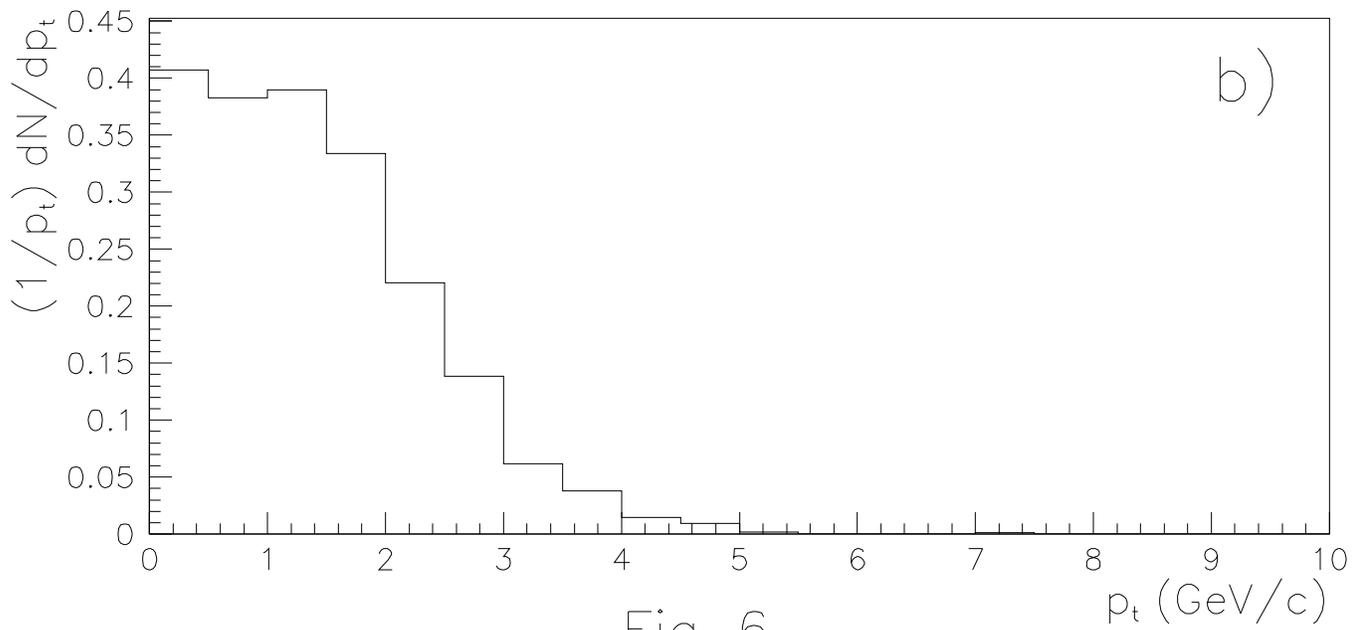
{\bf Fig. 6.} $p_t$--distributions (normalized to unity)  of beauty particles
in: a) $pp$ and b) $pW$
collisions at $\sqrt{s}=39$ GeV.

\end{document}